\documentclass[reprint,NumberedRefs]{JASAnew}

\usepackage{natbib}
\usepackage{graphicx,color,rotating}
\usepackage[latin1]{inputenc}
\usepackage{textcomp}
\usepackage{dcolumn}
\usepackage{bm}     
\usepackage{upgreek}
\usepackage{subdepth}  			
\usepackage{epstopdf}   		
\usepackage[latin1]{inputenc}
\usepackage{amsmath}
\usepackage{amssymb}
\usepackage{amsfonts}
\usepackage{array}
\newcolumntype{L}[1]{>{\raggedright\let\newline\\\arraybackslash\hspace{0pt}}m{#1}}
\newcolumntype{C}[1]{>{\centering\let\newline\\\arraybackslash\hspace{0pt}}m{#1}}
\newcolumntype{R}[1]{>{\raggedleft\let\newline\\\arraybackslash\hspace{0pt}}m{#1}}

\usepackage{xcolor}						
\definecolor{RED}{rgb}{1.00, 0.00, 0.00}
\definecolor{GREEN}{rgb}{0.00, 1.00, 0.00}
\definecolor{BLUE}{rgb}{0.00, 0.00, 1.00}
\definecolor{MAGENTA}{rgb}{1.00, 0.00, 1.00}

\newcommand{\upspace}{\rule{0ex}{3.0ex}}

\newcommand{\mr}[1]{\ensuremath{\mathrm{#1}}}
\newcommand{\myvec}[1]{\bm{#1}}
\newcommand{\ee}{\mathrm{e}}
\newcommand{\ii}{\mathrm{i}}
\newcommand{\dm}{\mathrm{d}}
\newcommand{\avr}[1]{\big\langle #1 \big\rangle}

\newcommand{\iot}{{\ii\omega t}}

\newcommand{\ve}{\varepsilon}
\newcommand{\veO}{\ve_0}

\newcommand{\pp}{\partial}

\newcommand{\nablabf}{\boldsymbol{\nabla}}

\newcommand{\Lapl}{\nabla^2}

\newcommand{\grad}{\nablabf}

\newcommand{\divop}{\nablabf\cdot}

\renewcommand{\etal}{\textit{et~al.}}

\newcommand{\DDD}{\myvec{D}}

\newcommand{\kc}{k_\mathrm{c}}

\newcommand{\ks}{k_\mathrm{s}}

\newcommand{\nnn}{\myvec{n}}

\newcommand{\rrr}{\myvec{r}}

\newcommand{\uuu}{\myvec{u}}

\newcommand{\vvv}{\myvec{v}}

\newcommand{\zerovec}{\boldsymbol{0}}

\newcommand{\cO}{c_0}

\newcommand{\Eac}{E_\mathrm{ac}}
\newcommand{\EacO}{E_\mathrm{ac}^0}

\newcommand{\etaB}{\eta^\mathrm{b}}

\newcommand{\etab}{\eta^\mathrm{b}}

\newcommand{\etaO}{\eta_0}

\newcommand{\Gamsl}{\Gamma_\mathrm{sl}}

\newcommand{\rhosl}{\rho_\mr{sl}}

\newcommand{\fres}{f_\mathrm{res}}

\newcommand{\kapO}{\kappa_0}

\newcommand{\rhoO}{\rho_0}

\newcommand{\SICel}{^\circ\!\textrm{C}}

\newcommand{\SIMHz}{\textrm{MHz}}

\newcommand{\SIkgm}{\textrm{kg}\:\textrm{m$^{-3}$}}
\newcommand{\SIkgpcm}{\SIkgm}
\newcommand{\SIm}{\textrm{m}}

\newcommand{\SImm}{\textrm{mm}}
\newcommand{\SImum}{\textrm{\textmu{}m}}

\newcommand{\SInm}{\textrm{nm}}

\newcommand{\SIpTPa}{\textrm{TPa}^{-1}}

\newcommand{\SIGPa}{\textrm{GPa}}
\newcommand{\SIPas}{\textrm{Pa}\:\textrm{s}}
\newcommand{\SImPas}{\textrm{mPa}\:\textrm{s}}

\newcommand{\SIs}{\textrm{s}}

\newcommand{\SImps}{\SIm\,\SIs^{-1}}

\newcommand{\nn}{\nonumber}
\newcommand{\beq}[1]{\begin{equation} \eqlab{#1}}
\newcommand{\eeq}{\end{equation}}
\newcommand{\bsub}{\begin{subequations}}
\newcommand{\esub}{\end{subequations}}
\def\bal#1\eal{\begin{align}#1\end{align}}
\def\balat#1#2\ealat{\begin{alignat}{#1} #2 \end{alignat}}
\def\bsubal#1 #2\esubal{\bsuba{#1}\begin{align}#2\end{align} \esuba}     
\def\bsubalat#1#2#3\esubalat{\bsuba{#1} \begin{alignat}{#2} #3 \end{alignat} \esuba}
\newcommand{\bsuba}[1]{\bsub \eqlab{#1}}
\newcommand{\esuba}{\esub}

\newcommand{\eqlab}[1]{\label{eq:#1}}
\renewcommand{\eqref}[1]{Eq.~(\ref{eq:#1})}
\newcommand{\eqnoref}[1]{(\ref{eq:#1})}

\newcommand{\figref}[1]{Fig.~\ref{fig:#1}}

\newcommand{\figlab}[1]{\label{fig:#1}}

\newcommand{\secref}[1]{Section~\ref{sec:#1}}

\newcommand{\seclab}[1]{\label{sec:#1}}
\newcommand{\tabref}[1]{Table~\ref{tab:#1}}
\newcommand{\tabsref}[2]{Tables~\ref{tab:#1} and~\ref{tab:#2}}
\newcommand{\tablab}[1]{\label{tab:#1}}

\newcommand{\sigmabf}{\bm{\sigma}}

\newcommand{\taubf}{\bm{\tau}}

\newcommand{\pare}[1]{\left( #1 \right)}
\newcommand{\subsc}[1]{_\mr{#1}}
\newcommand{\qqT}[1]{\quad \text{#1} \quad}
\newcommand{\dvisc}{\delta_\mathrm{visc}}

\newcommand{\abs}[1]{\left| #1 \right|}

\newcommand{\ttt}{\bm{t}}
\newcommand{\cLi}[1]{c_\mathrm{lo}^\mathrm{(#1)}}

\begin{document}

\title{Numerical study of the coupling layer between transducer and chip in acoustofluidic devices}

\author{William Naundrup Bod\'e}
\email{winabo@dtu.dk}
\affiliation{Department of Physics, Technical University of Denmark,\\ DTU Physics Building 309, DK-2800 Kongens Lyngby, Denmark}

\author{Henrik Bruus}
\email{bruus@fysik.dtu.dk}
\affiliation{Department of Physics, Technical University of Denmark,\\
DTU Physics Building 309, DK-2800 Kongens Lyngby, Denmark}

\date{28 January 2021}

\begin{abstract}
We study by numerical simulation in two and three dimensions the coupling layer between the transducer and the microfluidic chip in ultrasound acoustofluidic devices. The model includes the transducer with electrodes, the microfluidic chip with a liquid-filled microchannel, and the coupling layer between the transducer and the chip. We consider two commonly used coupling materials, solid epoxy glue and viscous glycerol, as well as two commonly used device types, glass capillary tubes and silicon-glass chips. We study how acoustic resonances in ideal devices without a coupling layer is either sustained or attenuated as a coupling layer of increasing thickness is inserted. We establish a simple criterion based on the phase of the acoustic wave for whether a given zero-layer resonance is sustained or attenuated by the addition of a coupling layer. Finally, we show that by controlling the thickness and the material, the coupling layer can be used as a design component for optimal and robust acoustofluidic resonances.
\end{abstract}
\maketitle

\section{Introduction}

The acoustic impedance matching techniques for piezoelectric sensors and transducers is in general a well-studied field, as exemplified by the recent review by Rathod.\citep{Rathod2020} However, specifically for ultrasound acoustofluidic devices, the role of the coupling layer (also known as the carrier or matching layer) between the transducer and the microfluidic chip remains poorly understood beyond one-dimensional (1D) planar systems.\citep{Hill2002, Glynne-Jones2012} Whereas the function of a  matching layer for 1D traveling waves through layered structures, is simply to couple acoustic energy more efficiently into subsequent layers, its role in resonant 1D acoustofluidic systems is less straightforward. As analyzed by Glynne-Jones, Boltryk, and Hill,\citep{Glynne-Jones2012} the function of the coupling layer may be more structural, or to isolate the transducer from the fluid layer. In the present work, we study the more complex case of a fully three-dimensional (3D) acoustofluidic system.

We present 3D numerical simulations of a piezoelectric transducer coupled to an acoustofluidic chip through a thin coupling layer consisting of either a solid glue or a viscous liquid. We consider two commonly used types of acoustofluidic devices sketched in \figref{systems}, glass capillary tubes and silicon-glass chips. Capillary-tube devices have been applied as acoustic particle traps relying on a small transducer that actuates a vertical resonant pressure mode locally in the capillary.\cite{Hammarstrom2012, Lei2013, Mishra2014, Gralinski2014} Silicon-glass devices have been applied for continuous-flow focusing and separation of particle suspensions relying on bulk actuation of horizontal resonance modes in embedded microchannels.\citep{Barnkob2010, Magnusson2017, Petersson2018}

\begin{figure}[t]
 \centering
 \includegraphics[width=\columnwidth]{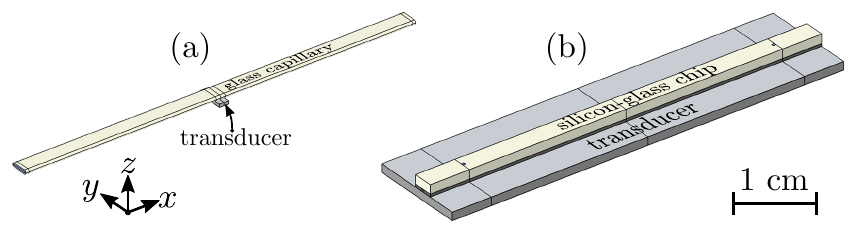}
 \caption[]{\figlab{systems} Sketch of the two types of acoustofluidic devices considered in this study. (a) A glass capillary (beige) mounted on a small piezoelectric transducer (gray). (b) A silicon-glass chip (black base, beige lid) mounted on a bulk piezoelectric transducer (gray). Both sketches are drawn to scale.}
\end{figure}

The paper is organized as follows: In \secref{theory}, we present the basic theory, including governing equations and boundary conditions. In \secref{capillary} we model the capillary-tube particle traps and show their dependency on the coupling-layer thickness. In \secref{bulk_devices} a similar analysis is carried out for the conventional acoustophoresis silicon-glass devices. Finally in \secref{discussion}, we present a concluding discussion regarding the criterion established in the two previous sections for designing acoustofluidic devices, with acoustic resonance modes that are relatively insensitive to the thickness of the coupling layer.

\section{Theory and model assumptions}
\seclab{theory}

In establishing the numerical model, we follow closely the theory presented by Skov \etal\citep{Skov2019} Our model consists of a lead-zirconate-titanate (PZT) piezoelectric transducer coupled to an elastic solid, which contains the fluid-filled microchannel. Theoretically, the system is described by three continuous fields: the electric potential $\varphi$ in the PZT transducer, the mechanical displacement field $\uuu$ in the elastic solid and the PZT transducer, and the acoustic pressure $p_1$ in the fluid. The coupling layer is described by $p_1$ if it is a liquid, and by $\uuu$ if it is a solid. Due to the linearity of the governing equations, all fields has a harmonic time dependence $\ee^{-\ii\omega t}$ with angular frequency $\omega = 2\pi f$ and frequency $f$. Thus, a given field has the structure $\tilde{f}(\rrr,t) = f(\rrr)\:\ee^{-\ii\omega t}$, and we need only determine the complex-valued space-dependent amplitude $f(\rrr)$.

The acoustic pressure field $p_1$ is modeled using the effective pressure acoustic theory by Bach and Bruus,\citep{Bach2018} where the viscous boundary layers are included analytically in the effective boundary conditions. Using the effective theory, the acoustic pressure $p_1$ in a fluid with density $\rhoO$, sound speed $\cO$, dynamic viscosity $\etaO$ and bulk viscosity $\etab$ is governed by the Helmholtz equation, and the acoustic velocity $\vvv_1$ is proportional to $\nablabf p_1$,
 \bsubalat{p1Gov}{2}
 \eqlab{Helmholtz}
 \Lapl p_1 &= -\kc^2 p_1, &
 \vvv_1 &= -\ii \frac{1-\ii\Gamma}{\omega \rhoO} \grad p_1,
 \\
 \eqlab{damping}
 \text{with } k_0 &= \frac{\omega}{\cO}, &
 \kc &=  \Big(1+\frac{\ii}{2}\Gamma \Big)\:k_0,
 \\
 \eqlab{v1}
 && \text{and }\;
 \Gamma &= \Big(\frac{\etaB}{\etaO }+\frac{4}{3}\Big) \frac{\omega\etaO  }{\rhoO \cO^2}.
 \esubalat
In cases where the fluid coupling layer thickness $\Delta$ is comparable or smaller than the viscous boundary-layer length scale $\dvisc=\sqrt{2\etaO/(\rhoO \omega)}$, the effective theory fails, and the full perturbation model is used instead.\citep{Muller2012, Muller2013}

The mechanical displacement field $\uuu$ is governed by the linear Cauchy equation involving the stress tensor $\sigmabf$,
 \beq{Cauchy}
 -\rhoO \omega^2  \uuu = \divop \sigmabf,
 \eeq
The components $\sigma_{ik}$ are related to the strain components $\frac{1}{2}(\partial_i u_k + \partial_k u_i)$ by the stiffness tensor $\bm{C}$, which for linear isotropic or cubic-symmetric elastic materials are written in the Voigt notation as
 \begin{equation}
 \resizebox{\columnwidth}{!}{$\displaystyle
 \eqlab{elastic_coupling}
 \pare{\begin{array}{c}
 \sigma_{xx} \\
 \sigma_{yy} \\
 \sigma_{zz} \\
 \hline
 \sigma_{yz} \\
 \sigma_{xz} \\
 \sigma_{xy}
 \end{array}}
 =
 \pare{
 \begin{array}{ccc|ccc}
 C_{11} & C_{12} & C_{12} & 0 & 0 & 0  \\
 C_{12} & C_{11} & C_{12} & 0 & 0 & 0  \\
 C_{12} & C_{12} & C_{11} & 0 & 0 & 0  \\
 \hline
 0 & 0 & 0 & C_{44} & 0 & 0 \\
 0 & 0 & 0 & 0 & C_{44} & 0 \\
 0 & 0 & 0 & 0 & 0 & C_{44} \\
 \end{array}
 }
 \,
 \pare{
 \begin{array}{c}
 \pp_x u_x \\
 \pp_y u_y \\
 \pp_z u_z \\
 \hline
 \pp_y u_z + \pp_z u_y \\
 \pp_x u_z + \pp_z u_x \\
 \pp_x u_y + \pp_y u_x
 \end{array}
 }.$}
 \end{equation}
Mechanical damping is implemented as complex-valued elastic moduli, defined as $C_{ik} = (1-\ii\Gamsl)c_{ik}$.
In the PZT transducer, the electric potential $\varphi$ is governed by the quasi-static Gauss equation involving the electric displacement $\DDD$,
 \beq{Gauss}
 \divop \DDD = 0.
 \eeq
Furthermore in PZT, the complete linear electromechanical coupling relating the stress and the electric displacement to the strain and the electric field is given by the Voigt notation as,
 \begin{equation}
 \resizebox{\columnwidth}{!}{$\displaystyle
 \eqlab{pzt_coupling}
 \pare{\begin{array}{c}
 \sigma_{xx} \\
 \sigma_{yy} \\
 \sigma_{zz} \\
 \hline
 \sigma_{yz} \\
 \sigma_{xz} \\
 \sigma_{xy} \\
 \hline
 D_x \\
 D_y \\
 D_z
 \end{array}}
 =
 \pare{
 \begin{array}{ccc|ccc|ccc}
 C_{11} & C_{12} & C_{13} & 0 & 0 & 0 & 0 & 0 & -e_{31} \\
 C_{12} & C_{11} & C_{13} & 0 & 0 & 0 & 0 & 0 & -e_{31} \\
 C_{13} & C_{13} & C_{33} & 0 & 0 & 0 & 0 & 0 & -e_{33} \\
 \hline
 0 & 0 & 0 & C_{44} & 0 & 0 & 0 & -e_{15} & 0\\
 0 & 0 & 0 & 0 & C_{44} & 0 & -e_{15} & 0 & 0\\
 0 & 0 & 0 & 0 & 0 & C_{66} & 0 & 0 & 0\\
 \hline
 0 & 0 & 0 & 0 & e_{15} & 0 & \epsilon_{11} & 0 & 0\\
 0 & 0 & 0 & e_{15} & 0 & 0 & 0 & \epsilon_{11} & 0\\
 e_{31} & e_{31} & e_{33} & 0 & 0 & 0 & 0 & 0 & \epsilon_{33}
 \end{array}
 }
 \,
 \pare{
 \begin{array}{c}
 \pp_x u_x \\
 \pp_y u_y \\
 \pp_z u_z \\
 \hline
 \pp_y u_z + \pp_z u_y \\
 \pp_x u_z + \pp_z u_x \\
 \pp_x u_y + \pp_y u_x \\
 \hline
 -\pp_x\varphi \\
 -\pp_y\varphi\\
 -\pp_z\varphi
 \end{array}
 }.$}
 \end{equation}

\subsection{Acoustic energy density}
Throughout this study, the time- and volume-averaged acoustic energy density $\Eac$ in the water-filled channel is used as a measure and indicator of how the acoustic resonances are affected by the coupling layer. In a fluid volume $V$ the averaged acoustic energy density is given as
 \beq{Eac}
 \Eac = \frac{1}{V} \int \Big(\frac{1}{4} \rhoO \abs{\vvv_1}^2 +\frac{1}{4}\kapO \abs{p_1}^2 \Big)\: \dm V.
\eeq

\subsection{Boundary conditions}
At the fluid-solid interface the boundary conditions are no-slip and continuous stress, together with zero stress on free surfaces. Introducing the mechanical displacement velocity $\vvv_\mr{sl}=-\ii \omega \uuu$  and a shear wave number $\ks = (1+\ii)/\dvisc$ the effective continuous velocity and stress boundary conditions become,\citep{Bach2018}
 \bsubal{BCeffective} \eqlab{no-slip}
 \nnn\cdot\vvv_1 &= \nnn\cdot\vvv_\mr{sl}
 + \frac{\ii}{\ks} \nablabf_\parallel\cdot(\vvv_\mr{sl}-\vvv_1),
 \\
 \eqlab{cont_stress}
 \sigmabf \cdot \nnn &= -p_1\nnn + \ii \ks \etaO \Big(\vvv_\mr{sl}+\frac{\ii}{\omega \rhoO}\grad p_1\Big),
 \esubal
where the unit vector $\nnn$ is the outward  surface normal from the solid domain. In experiments, the electrical signal is driven by a sine-wave function generator coupled to the transducer electrodes, this is implemented as a constant potential boundary condition on the electrode-transducer interface. Furthermore we assume no free charges, implemented as a zero flux condition on the electric displacement field. In the 3D capillary-tube device, symmetries are exploited such that the full system can be reduces to one quarter. The boundary conditions are listed in \tabref{tab_BCs}. Except for the symmetry conditions, the same boundary conditions applies for the silicon-glass device.

\begin{table}[t]
\centering
\caption{\tablab{tab_BCs} List of boundary conditions used in the modeled acoustofluidic systems. The unit vector $\nnn$ is the surface outward normal with respect to the solid domain, and $\ttt$ is any of the two tangential unit vectors.}
\begin{ruledtabular}
\begin{tabular}{ll}
Domain $\leftarrow$ boundary   &  Boundary condition  \\ \hline
Solid domain $\leftarrow$  air    & $\sigmabf\cdot \nnn = \zerovec$\\
Fluid domain $\leftarrow$  solid  &    \eqref{no-slip} \\
Solid domain $\leftarrow$  fluid   & \eqref{cont_stress}   \\
Fluid domain $\leftarrow$  air  & $p_1=0$ \\

PZT domain $\leftarrow$  bottom electrode  & $\varphi = 0$ \\

PZT domain $\leftarrow$  top electrode  & $\varphi = V_0$ \\

Solid domain $\leftarrow$  symmetry   &   $\uuu \cdot \nnn = 0$,  $\ttt\cdot \sigmabf \cdot \nnn = 0$ \\
Fluid domain $\leftarrow$  symmetry   & $\nnn \cdot \nablabf p_1 = 0$ \\
\end{tabular}
\end{ruledtabular}
\end{table}

\subsection{Unbounded perfectly matched layers}
For long systems like the capillary tubes, a no-reflection boundary condition can by established closer to the origin of the domain by using the perfectly matched layer (PML) technique, thus reducing the computational domain substantially. It involves a complex coordinate transformation of the form $x \to x + \frac{\ii}{\omega}\int^x \theta(x') \dm x'$, such that outgoing waves are attenuated within a distance comparable to the wavelength. The PML technique requires a choice of damping function $\theta$, the specific function is adopted from Berm\'{u}dez \etal\citep{Bermudez2007},
 \beq{PML}
 \theta(x) = \begin{cases} 0,
 & \text{for } x\leq L_\mr{cap},\\
 \frac{\beta}{L_\mr{pml} -(x-L_\mr{cap})} - \frac{\beta}{L_\mr{pml}},
 & \text{for } x> L_\mr{cap}.
 \end{cases}
 \eeq
The parameters defining $\theta(x)$ is chosen appropriately for a given system: $L_\mr{cap}$ is the position of the interface between the physical capillary tube and the PML domain, $L_\mr{pml}$ is the length of the PML domain, and $\beta$ is the damping strength. The axial coordinate $x$ is complex-valued for $x> L_\mr{cap}$ inside the PML domain. The function $\theta$ is classified as a continuous unbounded damping function, effective in terms of numerical error and reflections at the PML interface $x=L_\mr{cap}$.\citep{Bermudez2007}

\subsection{Numerical implementation in COMSOL Multiphysics}
The numerical model was implemented in the finite element software COMSOL Multiphysics\citep{Comsol55} using "Weak Form PDE" in the Mathematics module and closely following Ref.~\onlinecite{Skov2019}, where further implementation details can be found. The mesh settings is adopted from Ley and Bruus.\citep{Ley2017} The scripts were computed on a workstation with a 12 core 3.5 GHz CPU processor, and 128 GB ram.

\section{Capillary-tube particle traps}
\seclab{capillary}

As the first example, we investigate the capillary-tube device widely used as a versatile acoustic trap in many experimental studies.\cite{Hammarstrom2012, Lei2013, Mishra2014, Gralinski2014} The corresponding model system is sketched in \figref{capillary} indicating the different domains together with the PML layer and a zoom-in on the coupling layer. The dimensions and materials used in the numerical model are listed in \tabsref{capillary_dim}{material_values}, respectively. The model system is similar to the one studied by Ley and Bruus,\cite{Ley2017} but now the model is extended to include a PZT transducer and a coupling layer. Typically the capillary-tube device is characterized by having a standing half-wave-like resonance in the vertical direction. This is achieved with a PZT transducer having a predesigned mode at 5~MHz, also used in preliminary experiments by the Laurell group at Lund University.

\begin{figure}[t]
 \centering
 \includegraphics[width=\columnwidth]{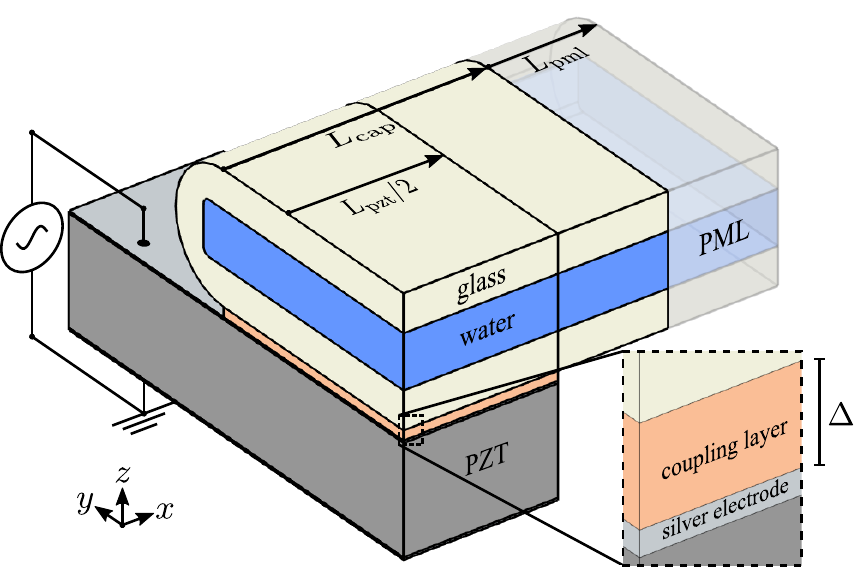}
 \caption[]{\figlab{capillary} One quarter of the capillary-tube-based model system with a zoom-in on the coupling layer of thickness $\Delta$. The model system includes a water-filled glass capillary coupled to a PZT transducer with silver electrodes. The top electrode is coupled to a time-harmonic function generator and the bottom electrode is grounded.}
\end{figure}

\begin{table}[t]
\centering
\caption{\tablab{capillary_dim} The length ($L$), width ($W$), and height ($H$) of the glass capillary tube (cap), the channel (ch), the  piezoelectric transducer (pzt), and the silver electrodes (el). The curvature of the outer and inner rounded corners are $240~\SImum$ and $25~\SImum$, respectively. The bottom $H_\mr{gl,bot}$ and top $H_\mr{gl,top}$ glass-wall thicknesses are both $140~\SImum$.}
\begin{ruledtabular}
\begin{tabular}{lrclr}
 Symbol       & Value         &        &  Symbol     & Value         \\  \hline
 $L_\mr{cap}$ & $1573~\SImum$  &        & $L_\mr{ch}$ & $1573~\SImum$  \\
 $W_\mr{cap}$ & $2280~\SImum$ &        & $W_\mr{ch}$ & $2000~\SImum$ \\
 $H_\mr{cap}$ & $480~\SImum$  & \qquad & $H_\mr{ch}$ & $200~\SImum$  \\
 $L_\mr{pzt}$ & $1160~\SImum$ &        & $L_\mr{el}$ & $1160~\SImum$ \\
 $W_\mr{pzt}$ & $3350~\SImum$ &        & $W_\mr{el}$ & $3350~\SImum$ \\
 $H_\mr{pzt}$ & $400~\SImum$  &        & $H_\mr{el}$ & $9~\SImum$    \\
\end{tabular}
\end{ruledtabular}
\end{table}

\begin{table}[h!]
\centering
\caption{\tablab{material_values} List of parameters used in the numerical simulations. Note that $C^{{}}_{12} = C^{{}}_{11} - 2 C^{{}}_{44}$ for isotropic materials. Isotropy in the $xy$-plane implies $C_{66}=\frac{1}{2}(C_{11}-C_{12})$ for the PZT.}
\begin{ruledtabular}
\begin{tabular}{lcrc}
 Parameter &  Symbol  & Value & Unit
 \\
 \hline
 \multicolumn{4}{l}{\textit{Water} at $25~\SICel$ \citep{Muller2014, Karlsen2016} \upspace} \\
 Mass density & $\rhoO$ & $997.05$ & $\SIkgpcm$ \\
 Speed of sound & $\cO$ & $1496.7$ & $\SImps$ \\
 Compressibility & $\kapO$ & $447.7$ & $\SIpTPa$ \\
 Dynamic viscosity & $\etaO$ & $0.890$ & $\SImPas$ \\
 Bulk viscosity & $\etab$ & $2.485$ & $\SImPas$
 \\

  \hline
 \multicolumn{4}{l}{\textit{Isotropic Pyrex borosilicate glass} \citep{CorningPyrex} \upspace}  \\
 Mass density & $\rho_\mr{sl}$ & $2230$ & $\SIkgpcm$ \\
 Elastic modulus & $c_{11}$ & $69.7$ & $\SIGPa$ \\
 Elastic modulus & $c_{44}$ & $26.2$ & $\SIGPa$ \\
 Mechanical damping coeff.        & $\Gamsl$ & 0.0004 & - \\
   \hline
 \multicolumn{4}{l}{\textit{Isotropic silver}\citep{AZOsilver2001} \upspace}  \\
 Mass density & $\rho_\mr{sl}$ & $10485$ & $\SIkgpcm$ \\
 Elastic modulus & $c_{11}$ & $133.9$ & $\SIGPa$ \\
 Elastic modulus & $c_{44}$ & $25.9$ & $\SIGPa$ \\
 Mechanical damping coeff.        & $\Gamsl$ & 0.0004 & - \\
   \hline
 \multicolumn{4}{l}{\textit{Cubic-symmetric silicon} \citep{Hopcroft2010} \upspace}  \\
 Mass density & $\rho_\mr{sl}$ & $2329$ & $\SIkgpcm$ \\
 Elastic modulus & $c_{11}$ & $165.7$ & $\SIGPa$ \\
 Elastic modulus & $c_{44}$ & $79.6$ & $\SIGPa$ \\
 Elastic modulus & $c_{12}$ & $63.9$ & $\SIGPa$ \\
 Mechanical damping coeff.        & $\Gamsl$ & 0.0001 & - \\
 \hline
 \multicolumn{4}{l}{\textit{Pz26 PZT ceramic} \citep{Ferroperm2017, Hahn2015} \upspace}  \\
 Mass density & $\rho_\mr{sl}$ & $7700$ & $\SIkgpcm$ \\
 Elastic modulus & $c_{11}$ & $168$ & $\SIGPa$ \\
 Elastic modulus & $c_{12}$ & $110$ & $\SIGPa$ \\
 Elastic modulus & $c_{13}$ & $99.9$ & $\SIGPa$ \\
 Elastic modulus & $c_{33}$ & $123$ & $\SIGPa$ \\
 Elastic modulus & $c_{44}$ & $30.1$ & $\SIGPa$ \\
 Coupling constant & $e_{15}$ & $9.86$ & C/m$^2$ \\
 Coupling constant & $e_{31}$ & $-2.8$ & C/m$^2$ \\
 Coupling constant & $e_{33}$ & $14.7$ & C/m$^2$ \\
 Electric permittivity & $\ve_{11}$ & $828$ & $\veO$ \\
 Electric permittivity & $\ve_{33}$ & $700$ & $\veO$ \\
 Mechanical damping coeff.        & $\Gamsl$ & 0.02 & - \\
\end{tabular}
\end{ruledtabular}
\end{table}

\subsection{The specific perfectly matched layer}
We implement no-reflection boundary conditions using a PML layer with parameter values $\beta = 2\cLi{gl}= 11.294$~m/s and $L_\mr{pml}=413~\SImum$. The superscript refer to the material, in this case glass (gl), and $\cLi{gl}$ and $\lambda_\mr{lo}^\mr{(gl)}$ is the longitudinal sound speed and wavelength, respectively. The numerical error introduced by using the PML is shown in \figref{PML}, in terms of the convergence parameter $C$, which for a given field solution $g$ is defined by
 \beq{PML_error}
 C(g) = \sqrt{\frac{\int \abs{g-g_\mr{ref}}^2 \dm V}{\int \abs{g_\mr{ref}}^2 \dm V}},
 \eeq
where $g_\mr{ref}$ is a reference solution. The integration domain is taken as the transducer region for $x\leq L_\mr{pzt}/2$ defined in \figref{capillary}. In \figref{PML}, the error measure $C$ is evaluated at six different geometries at a fixed frequency $f = 3.84\,\SIMHz$ with $\lambda_\mr{lo}^\mr{(gl)} = 1.47\,\SImm$. The reference solution is taken as $L_\mr{cap} = 1.35\lambda_\mr{lo}^\mr{(gl)}$. We choose the system length $L_\mr{cap} = 1.07\lambda_\mr{lo}^\mr{(gl)}$ such that the maximal numerical error due to the unbounded PML is estimated to be $C=3\times 10^{-4}$.

\begin{figure}[t]
 \centering
 \includegraphics[width=\columnwidth]{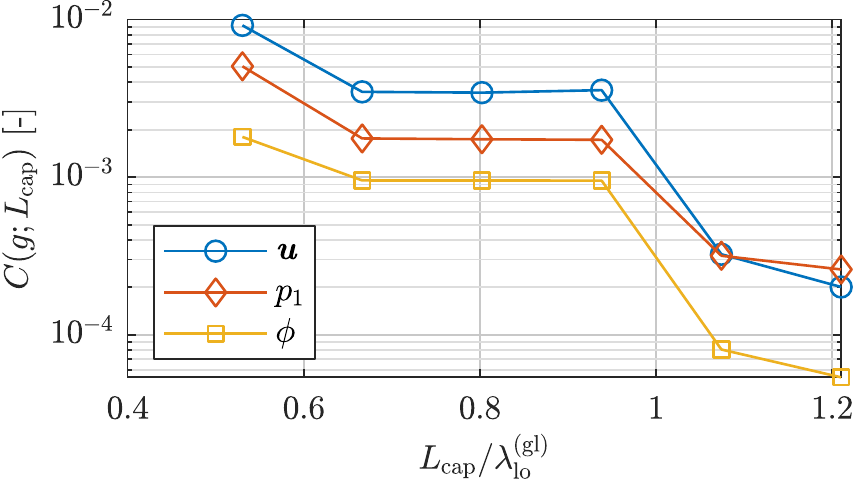}
 \caption[]{\figlab{PML} Numerical convergence $C$ for $\uuu$, $p_1$, and $\varphi$ in the PML of \figref{capillary} at frequency $\fres=3.84$~MHz with wavelength $\lambda_\mr{lo}^\mr{(gl)}=1.47\,\SImm$, PML length $L_\mr{pml}=413\,\SImum$, and the physical system length $L_\mr{cap}$ is varied as $L_\mr{cap}/\lambda_\mr{lo}^\mr{(gl)} = 0.53$, 0.67, 0.80, 0.94, 1.07,  and 1.21.}
\end{figure}

\subsection{Coupling-layer analysis in 3D and 1D}
\seclab{capilCoupling3D1D}

Continuing with the physics studies, we consider two coupling materials, a viscous mixture of 99 vol.\% glycerol and 1 vol.\% water, and a solid ED-20 epoxy resin, from now on referred to as glycerol and epoxy. The coupling material parameters used in the simulations are listed in \tabref{param_coupling}.

In practice, the glycerol coupling allows for reuseability of the acoustofluidic chip and/or the transducer, whereas the epoxy is used to ensure a well-defined but permanent coupling.\cite{Hawkes2001, Barnkob2010, Hammarstrom2010, Ohlsson2016, Bode2020} The effect of the coupling layer is investigated by calculating the resonances as a function of coupling-layer material and thickness $\Delta$ using the 3D model. For each coupling-layer thickness $\Delta$, the average acoustic energy density $\Eac$ in the water-filled channel is computed as a function of frequency from 3.0 to 4.5~MHz. Resonances are then identified as peaks in the acoustic energy spectrum $\Eac(f)$. The resonances are illustrated in the scatter plot of \figref{resonance_3D_and_1D}, where the points represent resonances at frequency $f$ for a coupling layer thickness $\Delta$ in the range from 0 to $100~\SImum$, with point areas proportional to $\Eac/\EacO$, where $\EacO$ is the acoustic energy density without a coupling layer. Also the resonances of the unloaded PZT resonances are plotted to indicate where the transducer is most active.

%
\begin{table}[t]
\centering
\caption{\tablab{param_coupling}  List of the coupling layer parameters for glycerol (a 99\% v/v glycerol and 1\% v/v water mixture) and epoxy at $20~\SICel$. The coefficient $C_{12}$ of the epoxy is obtained through the relation $C_{12} = C_{11} - 2C_{44}$.}
\begin{ruledtabular}
\begin{tabular}{lcrc}
 Parameter &  Symbol  & Value & Unit \\  \hline
 \textit{Glycerol} \citep{Slie1966, Negadi2017, Cheng2008} &  &  &  \\
 Mass density & $\rho\subsc{glc}$ & $1260.4$ & $\SIkgpcm$ \\
 Speed of sound & $c\subsc{glc}$ & $1922.8$ & $\SImps$ \\
 Compressibility & $\kappa\subsc{glc}$ & $214.6$ & $\SIpTPa$ \\
 Dynamic viscosity & $\eta\subsc{glc}$ & $1.137$ & $\SIPas$ \\
 Bulk viscosity & $\etab\subsc{glc}$ & $0.790$ & $\SIPas$ \\ \hline
 \textit{Epoxy}\citep{Perepechko1996} &  &  &  \\
 Mass density & $\rhosl$ & $1205$ & $\SIkgpcm$ \\
 Elastic modulus & $c_{11}$ & $9.583$ & $\SIGPa$ \\
 Elastic modulus & $c_{44}$ & $2.164$ & $\SIGPa$ \\
 Mechanical damping coeff. & $\Gamsl$ & $0.01$ & - \\
\end{tabular}
\end{ruledtabular}
\end{table}

\begin{figure}[t]
\centering
\includegraphics[width=\columnwidth]{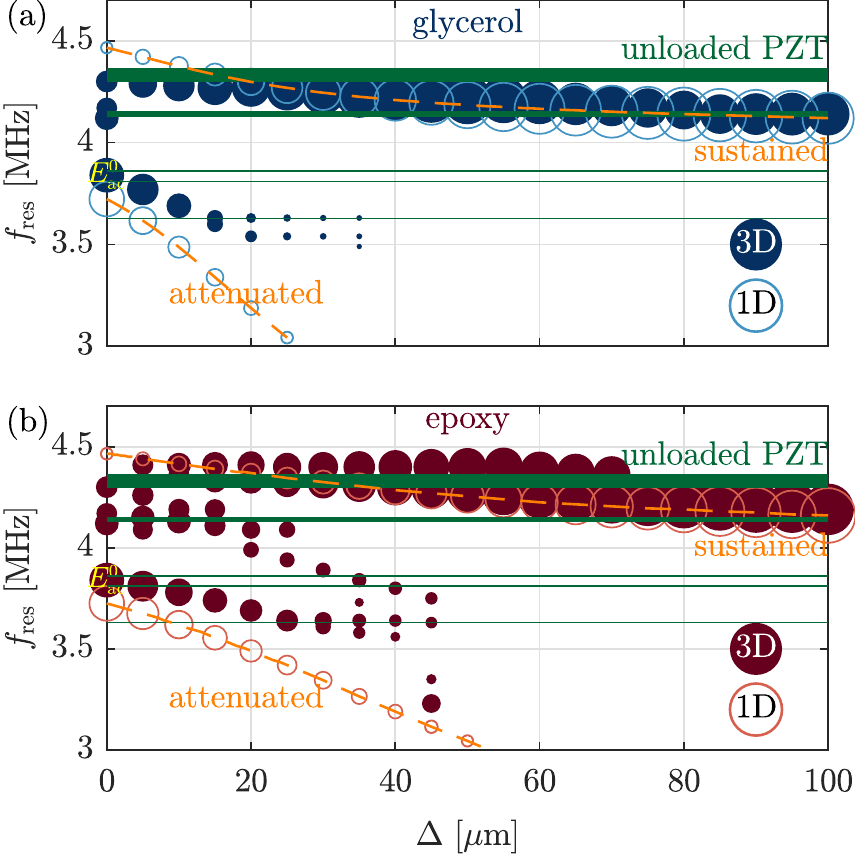}
\caption[]{\figlab{resonance_3D_and_1D} Resonance frequencies in the capillary-tube device, as a function of increasing coupling-layer thickness $\Delta$ for (a) glycerol and (b) epoxy. The 3D and 1D model resonance frequencies are plotted as filled and empty circles, respectively, with an area proportional $\Eac/\EacO$. The dashed lines represents the 1D resonance frequencies, indicating a sustained or attenuated behavior. The solid green lines indicate unloaded PZT resonances with a linewidth proportional to the logarithm of the acoustic energy density in the PZT.}
\end{figure}

The 3D model reveals a distinct behavior for both glycerol and epoxy coupling layers: As the coupling layer $\Delta$ increases, one resonance is attenuated ($\Eac$ decreases) and has a large downshift in frequency, whereas another is sustained ($\Eac$ increases) and has a small downshift in frequency. This behavior is also observed in an idealized 1D layer model along the vertical $z$-axis with seven domains: electrode, PZT, electrode, coupling layer, glass, water, glass of respective thicknesses $H_\mr{el}$, $H_\mr{pzt}$, $H_\mr{el}$,  $\Delta$, $H_\mr{gl,bot}$, $H_\mr{ch}$, and $H_\mr{gl,top}$. This 1D model takes into account only $z$-components, $z$-dependencies, densities and longitudinal sound speeds in the governing equations. The 1D model resonances are plotted together with the 3D resonances in \figref{resonance_3D_and_1D}. Of course, \figref{resonance_3D_and_1D} reveals that the 3D model exhibits more resonances than the 1D model due to the extended degrees of freedom in the transverse directions and the shear waves. However, for the two indicated modes in each system, the 1D model agrees fairly well with the 3D model, which indicates that the attenuated and sustained resonance effect can be explained by this 1D fluid-like model.

\begin{figure}[t]
    \centering
    \includegraphics[width=\columnwidth]{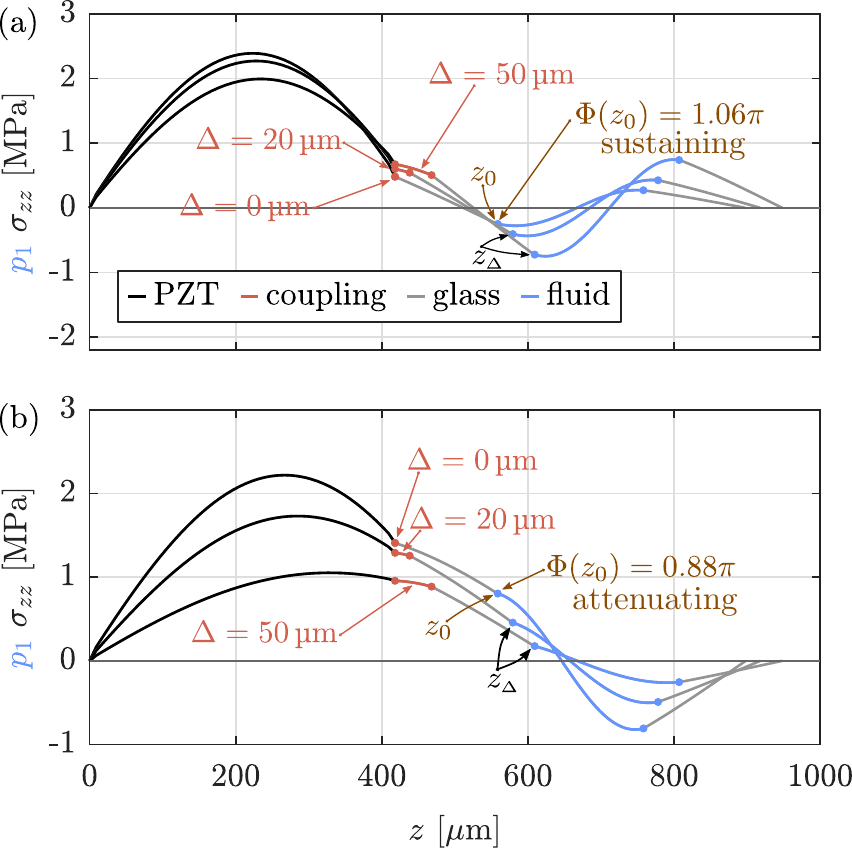}
    \caption[]{\figlab{phase_relation} Plots of normal stress $\sigma_{zz}(z)$ and pressure $p_1(z)$ in the 1D model with an epoxy coupling layer for the three layer thicknesses $\Delta=0$, 20, and $50~\SImum$ for (a) the sustained and (b) the attenuated zero-layer resonance modes. The profiles are plotted at a phase where the amplitude is at a maximum. The accumulated phase $\Phi$ is calculated without a coupling layer at the fluid-solid interface at $z_0=H_\mr{pzt}+2H_\mr{el}+H_\mr{gl,bot}$ (brown arrow).}
\end{figure}

\subsection{A design criterion for coupling layers}

The sustaining and attenuating behavior of the zero-layer resonances, as the layer thickness $\Delta$ is changed from 0 to $100~\SImum$, is elucidated by studying the pressure profiles for each of the resonances in the 1D model. For each profile we choose a temporal phase factor $\ee^{-\iot}$ that gives the maximum \textit{positive amplitude} in the PZT domain.

In \figref{phase_relation}(a) the 1D resonance pressure profiles are evaluated at three different epoxy coupling-layer thicknesses $\Delta = 0$, 20, and $50~\SImum$ for the sustained resonance. In \figref{phase_relation}(b), the same is shown for the attenuated resonance. Without the coupling layer, the fluid-solid interface is located at $z_0 = H_\mr{pzt}+ 2H_\mr{el} + H_\mr{gl,bot} = 558\,\SImum$, and with a coupling layer at $z_\Delta = z_0 + \Delta$. As a result, the value of the pressure $p_1(z_\Delta)$ is decreasing. In \figref{phase_relation}(a), labeled "sustained", we have $p_1(z_\Delta) < p_1(z_0) < 0$, yielding an \textit{increased magnitude} of the pressure as $\Delta$ is increased. In contrast, in \figref{phase_relation}(b), labeled "attenuated", we have $0 < p_1(z_\Delta) < p_1(z_0)$, yielding a \textit{decreasing magnitude} of the pressure for increasing $\Delta$. Clearly, if a given zero-layer resonance has a negative (positive) value of $p_1(z_0)$ for the specified temporal phase factor, the resonance is sustained (attenuated). The sign of $p_1(z_0)$ is determined by the accumulated spatial phase factor $\Phi(z_0)$ of the fluid-solid value $p_1(z_0)$ relative to the surface value $p_1(0)$. In the 1D model, $\Phi(z_0)$ is given by the wavenumber $k_i$ and layer thickness $H_i$ of each layer ($i =$ PZT, electrodes, and glass placed at  $z<z_0$). Consequently, we arrive at the criterion,
 \bsubal{criterion}
 \eqlab{sustainingRes}
 \text{sustaining coupling if }\; & \Phi(z_0) > \pi,
 \\
 \text{attenuating coupling if }\; & \Phi(z_0) < \pi,
 \\
 \eqlab{phase}
  \text{with}\;
 \Phi(z_0) = \sum_i k_i H_i &= \sum_i \frac{\omega}{\cLi{\mathit{i}}} H_i.
 \esubal
Note that this criterion is only valid for $|\Phi(z_0)| < \frac32\pi$. For the given capillary-tube device, the values of $\Phi(z_0)$  for the sustained and the attenuated zero-layer resonance is $3.32 = 1.06\pi$ and $2.77 = 0.88\pi$, respectively. This criterion is one of the main results of the paper. It can be used to design optimally coupled capillary devices with minimum attenuation caused by the coupling layer.

\subsection{Characteristic coupling-layer attenuation thickness $\bm{\Delta_0}$}

Based on the 1D model, we derive a semi-analytical estimate for the characteristic thickness $\Delta_0$, at which the acoustic energy is attenuated for the above-mentioned attenuated zero-layer resonance modes, see \figref{Delta0}. The pressure solution $p_{1,i}$ to the Helmholtz equation in each domain $i$, is written as
 \beq{sine-waves}
 p_{1,i} = p_{a,i}\sin(k_i z + \phi_i)\qqT{for} z \in \Omega_i.
 \eeq
At the interface between domain $i$ and $i+1$, the acoustic pressure and velocity must be continuous,
 \beq{1D_BCs}
 p_{1,i+1}=p_{1,i}, \qquad
 \frac{1}{\rho_{i+1}} \pp_z p_{1,i+1} =\frac{1}{\rho_{i}} \pp_z  p_{1,i}.
 \eeq
This results in an iterative formula for the amplitude $p_{a,i}$ and phase $\phi_i$ with coefficients $\beta_{i+1,i}$, $a_{i+1,i}$, and $b_{i+1,i}$,
 \bsubal{iteratinoEqu}
 \eqlab{paDef}
 p_{a,i+1} &=  \beta_{i+1,i}\: p_{a,i},
 \\
 \eqlab{phase-shift}
 \phi_{i+1} &= \phi_i - k_{i+1}\textstyle \sum^{i}_j H_j + \arctan(a_{i+1,i}/b_{i+1,i}),
 \\
 \eqlab{betaDef}
 \beta_{i+1,i} &= \sqrt{1+\cos^2(k_i \textstyle \sum^{i}_j H_j + \phi_i)(Z_{i+1,i}^2-1)},
 \\
 \eqlab{aDef}
 a_{i+1,i} &=\beta_{i+1,i}^{-1} \sin(k_{i}\textstyle \sum^{i}_j H_j+\phi_i),
 \\
 \eqlab{bDef}
 b_{i+1,i} &=\beta_{i+1,i}^{-1} Z_{i+1,i} \cos(k_{i}\textstyle \sum^{i}_j H_j+\phi_i),
 \\
 \eqlab{ZDef}
 Z_{i+1,i} &= \frac{\rho_{i+1}\cLi{\mathit{i}+1}}{\rho_{i}\cLi{\mathit{i}}}.
 \esubal
For a coupling layer (cl) made of either epoxy or glycerol, we have a mismatch $Z_\mr{cl,pzt} \ll 1$ in the acoustic impedance, and the pressure amplitude $p_1(z_0)$ at the  interface  $z_0$ can therefore be approximated as
 \bsubal{p1Approx}
 p_1(z_0) &\approx p_\mr{0} \sqrt{\alpha^2+\beta^2}\sin(k_\mr{cl} H_\mr{pzt})
 \\ \nn
 & \qquad \times \sin(\arctan(\alpha/\beta)+ k_\mr{gl} H_\mr{gl}) +\mathcal{O}(Z_\mr{cl,pzt}),
 \\
 \eqlab{a_and_b_Def}
 \text{with }\; & \alpha  = Z_\mr{gl,cl}\sin(k_\mr{cl}\Delta) \;\text{ and }\;
 \beta = \cos(k_\mr{cl} \Delta).
 \esubal
\begin{figure}[b]
    \centering
    \includegraphics[width=\columnwidth]{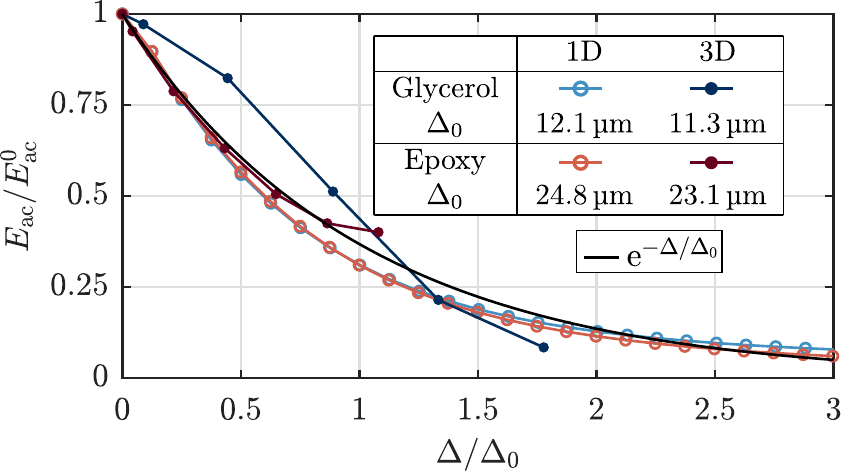}
    \caption[]{\figlab{Delta0} The normalized acoustic energy density $\Eac/\EacO$ for the attenuated capillary-tube resonance versus the normalized coupling-layer thickness $\Delta/\Delta_0$ for glycerol and epoxy, simulated in the 3D and in the 1D model. The exponential function $\ee^{-\Delta/\Delta_0}$ (black) is inserted as a guide to the eye.}
\end{figure}
By further assuming $\Eac\propto  p_1^2(z_0)$ and $k^2_\mr{cl}\Delta^2 \ll 1$, the lowest order functional dependence in the layer-thickness $\Delta$ becomes
 \bal
 \Eac &\propto  p_\mr{0}^2 \sin^2(k_\mr{cl}H_\mr{pzt})
 \\ \nn
 & \qquad \times\Big[\cos^2(k_\mr{gl} H_\mr{gl})
 - Z_\mr{gl,cl} k_\mr{cl}\Delta \sin(2k_\mr{gl} H_\mr{gl})\Big].
 \eal
By setting $\Eac = 0$, we extract the characteristic thickness scale $\Delta_0$, at which the resonance is attenuated
 \beq{Delta0}
 \Delta_0 = \frac{Z_\mr{cl,gl} \cLi{cl}}{2\omega} \cot\!\bigg( \frac{\omega H_\mr{gl}}{\cLi{gl}} \bigg).
 \eeq
In \figref{Delta0} the normalized acoustic energy density $\Eac(\Delta)/\EacO$ is plotted as a function of the normalized coupling-layer thickness $\Delta/\Delta_0$. The acoustic energy is seen to be attenuated on the length scale $\Delta_0$ predicted by \eqref{Delta0}, with $\Delta_0 = 12.1\,\SImum$ and $24.8\,\SImum$ for glycerol and epoxy, respectively. For both coupling materials, we observe an approximate exponential decay,  $\Eac(\Delta)/\EacO = \ee^{-\Delta/\Delta_0}$. The figure also show, that the attenuation computed in the 3D model is captured fairly well by the 1D model, including good quantitative agreement within 7\% between the estimated characteristic length scale $\Delta_0$ in 1D and 3D, respectively.

\section{Acoustophoretic bulk devices}
\seclab{bulk_devices}

We now move on to the second type of acoustofluidic devices, namely the bulk silicon-glass devices used in many lab-on-a-chip applications, as reviewed by Lenshof, Magnusson and Laurell.\citep{Lenshof2012a} As sketched in \figref{systems}(b), these devices consist of a silicon-glass-based acoustofluidic chip coupled to a bulk PZT transducer. In contrast to the capillary-tube devices, the manipulation of particles in the silicon-glass chip relies on horizontal half-wave pressure resonances. Because the pressure half-wave is anti-symmetric around the vertical center plane of the channel, the symmetric motion actuated by a usual PZT transducer must be broken. This is normally done geometrically, by placing the transducer off-center,\citep{Barnkob2010, Muller2013, Tahmasebipour2020} or by splitting the top electrode of the transducer, and actuate it by an anti-symmetric voltage actuation.\citep{Moiseyenko2019, Bode2020} In this work we use the former method, and displace the silicon-glass chip by $y_0 = 1\,\SImm$ with respect to the $(xz)$-mirror-plane of the PZT transducer, see \figref{acoustophoresis}.

\subsection{Coupling layer analysis in 2D}
\seclab{SiGlassCoupling2D}

It is well known that even with an ideal long straight channel in such devices, axial variations and acoustic hot spots appears along the channel.\citep{Augustsson2011} However, when studying a region in the channel near a local maximum in the acoustic field, where the axial gradients are vanishing small, 2D models describes the acoustic fields very well.\citep{Augustsson2011, Muller2013} In this analysis, we therefore study the silicon-glass device in a 2D model as shown in \figref{acoustophoresis}. As introduced in \secref{theory}, the PZT transducer is a $z$-polarized Meggitt-Pz26 transducer, but now with a thickness of $1000~\SImum$ with a resonance mode near 2~MHz. The materials and dimensions used in the silicon-glass-device simulations are listed in \tabsref{material_values}{silicon_glass_dim}, respectively.

\begin{figure}[t]
    \centering
    \includegraphics[width=\columnwidth]{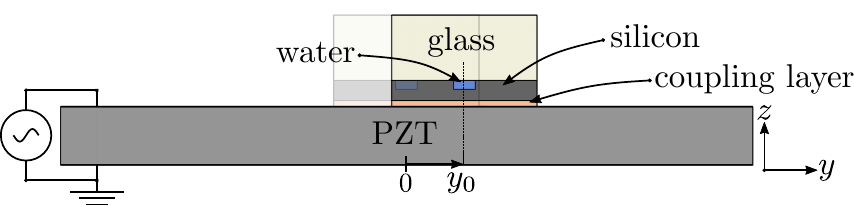}
    \caption[]{\figlab{acoustophoresis} Cross-section in the vertical $yz$-plane of the silicon-glass device in \figref{systems}(b) showing the PZT transducer, the coupling layer, and the silicon-glass chip with the microchannel. The chip is displaced from the PZT center plane by $y_0 = 1$~mm. The sketch defines the 2D model.}
\end{figure}

\begin{table}[t]
\centering
\caption{\tablab{silicon_glass_dim} The width ($W$) and height ($H$) of the silicon base (si), the
glass cover (gl), the channel (ch), the  piezoelectric transducer (pzt), and the silver electrodes (el).}
\begin{ruledtabular}
\begin{tabular}{lrclr}
 Symbol            & Value     &        &  Symbol       & Value \\  \hline
 $W_\mr{si}$  & $2520~\SImum$  &        & $H_\mr{si}$   & $350~\SImum$ \\
 $W_\mr{gl}$  & $2520~\SImum$  &        & $H_\mr{gl}$   & $1130~\SImum$ \\
 $W_\mr{ch}$  & $377~\SImum$   & \qquad & $H_\mr{ch}$   & $157~\SImum$ \\
 $W_\mr{pzt}$ & $12000~\SImum$ &        & $H_\mr{pzt}$  & $982~\SImum$ \\
 $W_\mr{el}$  & $12000~\SImum$ &        & $H_\mr{el}$   & $9~\SImum$ \\
\end{tabular}
\end{ruledtabular}
\end{table}

The coupling layer analysis is analogous  to the one in  \secref{capillary}. The acoustic resonances are located as peaks in the acoustic energy density spectrum $\Eac(f)$ in the frequency range 1.5 to 2.5~MHz as a function of coupling material (glycerol or epoxy) and layer thickness $\Delta$ from  0 to 100~\SImum.

Similar to \figref{resonance_3D_and_1D}, we show in \figref{SiGlass_resonances}(a) a scatter plot, where the points represent resonances at frequency $f$ for a coupling layer thickness $\Delta$ in the range from 0 to $100~\SImum$, with point areas proportional to $\Eac/\EacO$. Multiple resonances are identified, however, the one at $\fres=1.940~\SIMHz$ stands out with $\EacO$ being more than 60 times larger than any other zero-layer peak. The frequency of this resonance is nearly independent of the coupling material and the layer thickness $\Delta$. However, in \figref{SiGlass_resonances}(b) we see a fundamental difference between the two coupling materials: for glycerol the normalized acoustic energy density $\Eac/\EacO$ decreases rapidly to nearly zero at a length scale ${\sim}100\,\SInm$, whereas for epoxy $\Eac/\EacO$ stays nearly constant up to $\Delta \sim 1\,\SImum$ followed by a slow drop to 0.75 at $\Delta = 10\,\SImum$  and 0.2 at  $\Delta \sim 50\,\SImum$. This behavior may be explained by the geometry of the acoustics and the fundamental mechanical difference between elastic solids and viscous fluids. In the silicon-glass device, the direction of the standing pressure half-wave is orthogonal to the transducer polarization, and to excite this resonance mode, the transmission of shear-waves from the transducer to the microchannel is required. However, only a solid coupling layer, and not a viscous fluid, can support such transmission of shear waves.

\begin{figure}[t]
    \centering
    \includegraphics[width=\columnwidth]{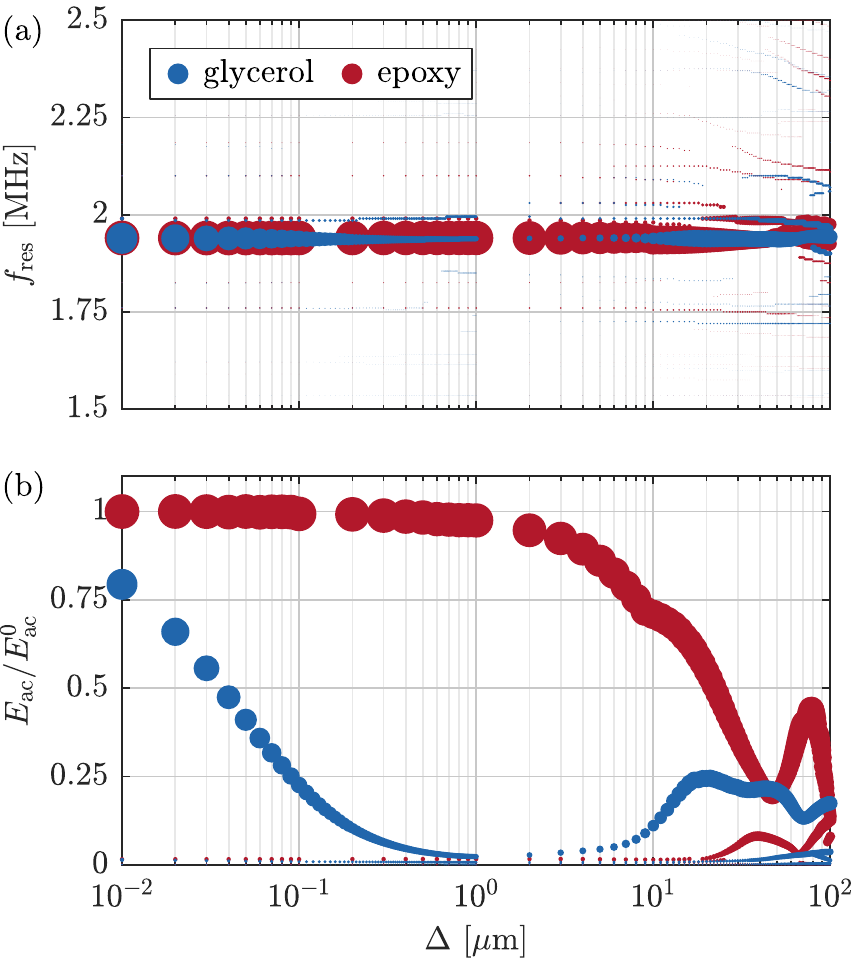}
    \caption{\figlab{SiGlass_resonances} Scatter plots with point areas proportional to $\Eac/\EacO$
    of responses to increasing coupling-layer thickness $\Delta$ simulated in the 2D model of the silicon-glass device. (a)~Resonance frequencies $f_\mr{res}$. (b) Normalized acoustic energy density $\Eac/\EacO$.}
\end{figure}

\subsection{Dissipation in the glycerol coupling layer}
\seclab{DissipationGlCoupling}

The critical glycerol coupling-layer thickness $\Delta_\mr{crit} = 100$~nm observed in \figref{SiGlass_resonances}(c) requires a physical explanation, as this length scales is far from any of the geometrical sizes or acoustic wavelengths in the system. The small thickness $\Delta < 0.1$~mm of the coupling layer implies large shear strain rates, and a large amount of viscous dissipation. We assume that $\Delta_\mr{crit}$ is the coupling-layer thickness where the time-averaged viscous dissipation power $\avr{P_\mr{crit}^\mr{visc}}$ in the glycerol coupling layer equals the time-averaged acoustic power $\avr{P_\mr{ac}}$ delivered to the half-wave pressure  resonator. For an ideal resonator with height $H_\mr{ch}$, average acoustic energy density $\Eac$, and quality factor $Q$, together with a coupling layer of thickness $\Delta_\mr{crit}$, width $W_\mr{glc}$, and dynamic viscosity $\eta_\mr{glc}$, we obtain
 \bsub
 \eqlab{PowerDissipation}
 \bal
 \eqlab{equalPower}
 \avr{P_\mr{ac}} &= \avr{P_\mr{crit}^\mr{visc}},
 \\
 \eqlab{PacDef}
 \avr{P_\mr{ac}} &= \int \avr{p_1 \vvv_1\cdot\nnn}\: \dm A
 = \frac{16\pi}{Q} \cO \Eac H_\mr{ch},
 \eal
 \bal
 \eqlab{PviscDef}
 \avr{P_\mr{crit}^\mr{visc}} &= \int \avr{\nablabf\vvv_1 : \taubf}\: \dm V
 \approx \frac{4 \pi^2 \eta_\mr{glc} \Eac }{Q^2 \rhoO \Delta_\mr{crit}}W_\mr{glc}.
 \eal
 \esub
Solving for $\Delta_\mr{crit}$, we obtain
 \beq{Delta_crit}
 \Delta_\mr{crit} = \frac{\pi \eta_\mr{glc} W_\mr{glc} }{4Q \rhoO \cO H_\mr{ch} }.
 \eeq
The effect of the surrounding silicon-glass chip is included in the quality factor $Q = f_\mr{res}/\Delta f$, found from the resulting full-width $\Delta f$ at half maximum of the corresponding resonance peak $\Eac(f)$ at the resonance frequency $f_\mr{res}$. The estimate for $\Delta_\mr{crit}$ is validated numerically by varying the material and geometrical parameters in \eqnoref{Delta_crit}. The chosen material and geometric variations are listed in \tabref{delta_crit} together with critical thicknesses and quality factors. Using the 2D model, the acoustic energy density $\Eac$ is simulated, and the result is normalized by $\EacO$, the value without a coupling layer. In \figref{delta_crit} the simulated $\Eac/\EacO$ is plotted versus the normalized coupling-layer thickness  $\Delta/\Delta_\mr{crit}$, and for the wide range of parameters, it is seen that indeed $\Eac/\EacO$ decays on the critical coupling-layer thickness scale $\Delta_\mr{crit}$.

\begin{table}[t]
\centering
\caption{Critical coupling-layer thickness $\Delta_\mr{crit}$ and quality factor $Q$ in nine different system configurations, categorized as material and geometry variations.}
\label{tab:delta_crit}
\resizebox{\columnwidth}{!}{%
\begin{tabular}{c|ccccccc|cc}
\hline\hline
 & \multicolumn{7}{c|}{Material}                 & \multicolumn{2}{c}{Geometry} \\ \hline
Parameter &
  $0.05\eta_\mr{glc}$ &
  $0.1\eta_\mr{glc}$ &
  $0.2\eta_\mr{glc}$ &
  $0.5\eta_\mr{glc}$ &
  $\eta_\mr{glc}$ &
  $2\eta_\mr{glc}$ &
  $2\rho_0$ &
  $0.5H_\mr{ch}$ &
  $0.5W_\mr{glc}$ \\
$\Delta_\mr{crit}$ [nm] & 3.4 & 6.9 & 13.7 & 34.3 & 68.5 & 137.0 & 73.0 & 202.2           & 38.5           \\
$Q$ [-]            & 140   & 140   & 140    & 140    & 140    & 140     & 140    & 95               & 125              \\ \hline\hline
\end{tabular}%
}
\end{table}

\begin{figure}[t]
    \centering
    \includegraphics[width=\columnwidth]{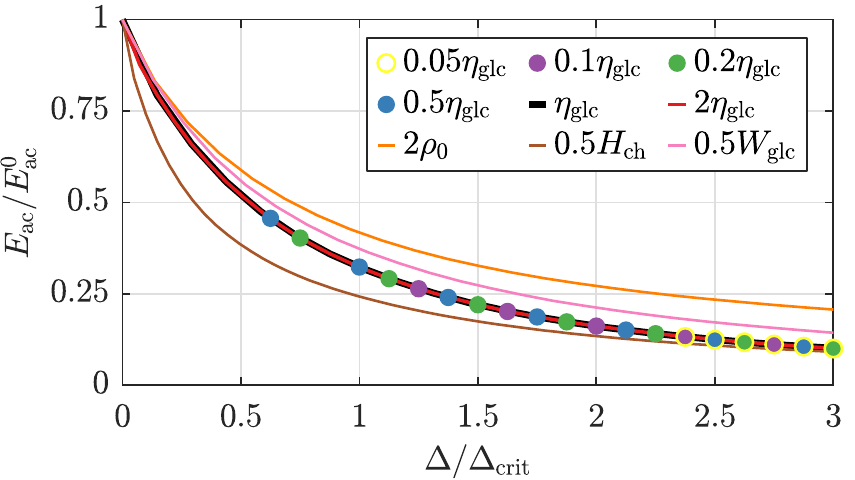}
    \caption[]{\figlab{delta_crit} Simulated normalized acoustic energy density $\Eac/\EacO$ versus the normalized coupling layer thickness $\Delta/\Delta_\mr{crit}$ for the nine different system configurations listed in \tabref{delta_crit}.}
\end{figure}

\section{Conclusion}
\seclab{discussion}

We have developed a numerical 3D model to study the role of coupling layers in acoustofluidic devices.  The model includes the PZT transducer with electrodes, the coupling layer, and the acoustofluidic chip with the fluid-filled microchannel. The model is used to study two well-known types of acoustofluidic devices: a glass capillary tube and a silicon-glass chip, classified as vertical and horizontal resonators, respectively, relative to the polarization axis of the transducer. For each device, a viscous glycerol and a solid epoxy coupling layer was studied.

For vertical resonators, the capillary-tube device, we have found that for a given zero-layer resonance, the coupling layer can either result in a sustaining or attenuating resonance. We have established the criterion~\eqnoref{criterion} to predict which of the two behaviors will occur, based on a relation involving the phase of the acoustic wave. For the attenuated waves, we have derived expression~\eqnoref{Delta0} to estimate the characteristic layer thickness $\Delta_0$ at which the resonance is attenuated.

For horizontal resonators, the silicon-glass device, the acoustic resonances are partially powered by shear-wave transmission through the coupling layer. Since a fluid cannot sustain such a shear, the glycerol coupling layer works as a dissipative layer. A critical viscous dissipation thickness $\Delta_\mr{crit}$ was presented in \eqref{Delta_crit}, based on scaling arguments in a 1D two-component model.

The simulation results presented, has led to the formulation of design rules for choosing an optimal coupling layer between the piezoelectric transducer and the acoustofluidic device. The design rules involves material parameters, geometrical parameters, and information about the orientation of the given acoustic resonance mode relative to the polarization axis of the transducer. We hope that these rules will prove useful, and that their limitations will be understood better by experimental validation.


\section*{Acknowledgments}
The authors were supported by Independent Research Fund Denmark, Technology and Production
Sciences (Grant No. 8022-00285B).

%
%


\end{document}